\documentclass[sn-mathphys]{sn-jnl}
\usepackage{graphicx}
\usepackage{amsmath, amssymb}

\usepackage{multirow} 
\usepackage{array,booktabs}

\usepackage{algpseudocode}

\newcommand{\sset}[1]{\mathcal{#1}}

\begin{document}

\title{Scalable almost-linear dynamical Ising machines}

\author{\fnm{Aditya} \sur{Shukla}}
\author{\fnm{Mikhail} \sur{Erementchouk}}
\author{\fnm{Pinaki} \sur{Mazumder}}
\affil{\orgdiv{Department of Electrical Engineering and Computer Science},
  \orgname{University of Michigan}, \orgaddress{\city{Ann Arbor},
    \postcode{48104}, \state{MI}, \country{USA}}}

\abstract{
  The past decade has seen the emergence of Ising machines targeting hard
  combinatorial optimization problems by minimizing the Ising
  Hamiltonian with spins represented by continuous dynamical variables.
  However, capabilities of these machines at larger scales are yet to be fully
  explored. We investigate an Ising machine
  based on a network of almost-linearly coupled analog spins. We show
  that such networks leverage the computational resource similar to that of
  the semidefinite positive relaxation of the Ising model. We estimate the
  expected performance of the almost-linear machine and benchmark it on a
  set of $\left\{0, 1\right\}$-weighted graphs. We show that the running
  time of the investigated machine scales polynomially (linearly
  with the number of edges in the connectivity graph). As an example of the
  physical realization of the machine, we present a CMOS-compatible
  implementation comprising an array of vertices efficiently storing the
  continuous spins on charged capacitors and communicating externally via
  analog current.
}
  

\maketitle

\section{Introduction}

Many existing practical optimization problems such as resource allocation,
traffic control, cell placement and interconnection routing within in a
very large-scale integrated (VLSI) chip, as well as emerging problems like
polymer modelling~\cite{Babbush2014ConstructionAnnealing}, protein
folding~\cite{Fraenkel1993ComplexityFolding} and medical image
segmenting~\cite{ref_imag} are NP hard, that is, as the problem scales in
size, they require exponentially more computing resources to obtain an
exact solution.
As the growth of the problem size outpaces scaling of commercial
processors, alternative faster computing techniques are being actively
sought.

One such alternative is based on classical spin systems on graphs. The
computational significance of these systems lies in a tight connection
between the distribution of spins delivering the lowest energy and the
maximal cut of the graph~\cite{barahona_computational_1982}. In turn, the
maximal cut problem is NP-complete~\cite{miller_reducibility_1972,
  gareySimplified1976} and therefore finding the ground state of the Ising
model can be employed for solving other NP-hard problems. This principle
was explicated in~\cite{Lucas2014IsingProblems}, where it was shown that
all Karp's original NP-complete problems can be solved by finding the
ground state of Ising models with specially crafted Hamiltonians.

The computational capabilities of spin systems are known for several
decades. Based on analogy with the thermal relaxation to the ground state,
a well-known class of algorithms with wide area of applications, simulated
annealing, was invented in~\cite{Kirkpatrick1983OptimizationAnnealing}.
The success of simulated annealing motivated several digital
annealing processors~\cite{Yamaoka201520k-spinAnnealing,
  Takemoto2020AProblems, su312020, takemoto144Kb2021,
  okuyamaComputing2016, Yamamoto2017AFPGAs}, which utilized the
direct analogy between binary states of classical spins in the Ising model
and binary devices. To emulate the thermal effect helping to traverse the
configuration space of the spin system, randomized spin-flips were
implemented using on-chip random bit-sequence generators.

Recently, a new generation of Ising machines based on continuous dynamics
has emerged~\cite{Ahmed2021AProblems, Bashar2020ExperimentalProblem,
  Parihar2017VertexNetworks, Raychowdhury2019ComputingSystems,
  Marandi2014NetworkMachine, McMahon2016AConnections, Bohm2019AProblems,
  Leleu2017CombinatorialSystems, Molnar2018APerformance,
  afoakwaBRIM2021}. These machines do not attempt to represent the
binary states of the classical spin. Instead, they leverage the emergent
capability of selected dynamical systems to deliver the ground state of the
Ising Hamiltonian.

The continuous dynamical Ising machines are demonstrated the ability to achieve good
quality solutions on benchmark problems, such as \texttt{GSet}~\cite{gset}.
At the same time, the problem of scaling of the computational budget, for
instance, running time, drew much less attention. The scaling problem is
important in the context of NP-complete problem, especially considering the
APX-hardness of the max-cut problem~\cite{papadimitriou_optimization_1991,
  khot_optimal_2004}. This means that the approximation achievable in time
scaling polynomially with the problem size is limited if
$\mathrm{P} \ne \mathrm{NP}$. The existing data on scaling of dynamical Ising
machines~\cite{hamerlyExperimental2019, leleuScaling2021a} demonstrate
super-polynomial scaling: the best result reported
in~\cite{leleuScaling2021a} is $\mathcal{O}\left(e^{\sqrt{N}} \right)$, where
$N$ is the number of graph nodes. Such scaling effectively puts large
NP-hard problem out of the reach.

In~\cite{erementchoukComputational2022}, we have shown that Ising machines
based on oscillator networks realizing the Kuramoto model of
synchronization can demonstrate the polynomial scaling, while delivering
solutions of good quality. This is related to the fact that the equations
of motion of the Kuramoto model in the synchronized regime implement the
gradient-descent solution of the rank-2 semidefinite programming
relaxation. It should be noted in this regard that large scale physical
realizations of Ising machines must resolve fundamental challenges
associated with the high degree of integration. The dynamical model
governing the Ising machine must be simple enough to admit a
straightforward realization and robust to withstand unavoidable
imperfections and variations. At the same time, this must be achieved
without sacrificing computational capabilities.

In this work, we investigate an approach of realizing dynamical Ising machines
that overcomes the challenges associated with oscillatory systems while
retaining their computational capabilities. We consider an almost-linear
dissipative dynamical system on a graph. We show that this model presents an
approximation of the rank-$2$ semidefinite positive (SDP) relaxation of the
Ising model. As a result, the model is capable of providing a good
approximation in time, which scales polynomially with the number of edges.
We show that the machine is characterized by the integrality gap close to
that of the SDP relaxation and demonstrates the core performance comparable
with the state-of-the-art solves.

As a proof-of-concept, we present a CMOS-compatible combinatorial optimizer
that implements the investigated almost-linear Ising machine. The
analog-digital mixed mode computing system is based on fully integrated
components using a 130 nm CMOS technology.

This paper is organized as follows. In Section 2, we present the necessary
theoretical background and find the integrality gap of the almost-linear
Ising machine. In Section 3, we consider a software simulation of the
machine, and investigate its performance on benchmark tests and its scaling
properties. In Section 4, we describe the CMOS computing system
implementing the almost-linear dynamical Ising machine.

\section{Dynamical Ising machines}

\subsection{Ising model and NP-hard problems}

Let $\sset{G} = \{\sset{V}, \sset{E}\} $ be a graph with $N$ vertices and
$M$ edges given by sets $\sset{V}$ and $\sset{E}$, respectively. The Ising
model on $\sset{G}$ deals with binary (taking values $\pm 1$) functions on
the set of graph vertices, $\boldsymbol{\sigma} : \sset{V} \to \{-1, 1\}$.
Alternatively, such functions can be understood as assigning a binary
variable $\sigma_i$ to the $i$-th vertex resulting in configuration
$\boldsymbol{\sigma} = \{\sigma_i\}$. The sets of nodes where $\boldsymbol{\sigma}$ takes positive and
negative values define partitioning of the graph nodes
$\sset{V} = \sset{V}_+ \cup \sset{V}_-$. Finding the partitioning with the
maximal number of edges connecting nodes in $\sset{V}_+$ and $\sset{V}_-$
is the maximal cut problem. This problem is
NP-complete~\cite{miller_reducibility_1972, gareySimplified1976} and,
therefore, finding its solution provides ways to solve other NP-complete
problems~\cite{Lucas2014IsingProblems}.

The Ising model is defined by assigning the energy to configurations
\begin{equation}
  {H}_I (\boldsymbol{\sigma})= \frac{1}{2} \sum_{m,n=1}^{N} A_{m,n} \sigma_m \sigma_n,
\label{eq_isingmod}
\end{equation}
where $A_{m,n}$ is the graph adjacency matrix. It is straightforward to
show that finding the ground state ($\boldsymbol{\sigma}$ yielding the lowest
${H}_I (\boldsymbol{\sigma})$) of the Ising model is equivalent to solving the
max-cut problem~\cite{barahona_computational_1982}. The function taking
values $1$ on edges connecting $\sset{V}_+$ and $\sset{V}_-$ and $0$
elsewhere can be written in terms of $\boldsymbol{\sigma}$ as
$(1 - \sigma_m \sigma_n )/2$. Thus, the total number of cut edges is
\begin{equation}\label{eq:max-cut-ham}
 C_{\sset{G}} \left( \boldsymbol{\sigma} \right) 
            = \frac{1}{4} \sum_{m,n} A_{m,n} \left(1 - \sigma_m \sigma_n \right)
    = \frac{M}{2} - \frac{1}{2} {H}_I\left(\boldsymbol{\sigma}\right).
\end{equation}

\subsection{Ising machines based on the Kuramoto model}

The operational principles of the presented architecture are based on a
proper adaptation of the principles enabling Ising machines based on
synchronizing oscillator networks. Therefore, we first review the
computational capabilities of such machines, and then we abstract them from
the oscillatory dynamics to formulate the dynamical model of our main
interest.

The Kuramoto model was originally introduced to study the synchronization
phenomenon in networks with inhomogeneous natural frequencies. However, in the
computing context, where the informational content is associated with
relative phases of the oscillators, the frequency variation can be expected
to play a detrimental role. Therefore, we consider a network of oscillators
with identical natural frequencies. In this case, the relative phases obey
the equations of motion
\begin{equation}\label{eq:qkm-eqm}
        \frac{1}{K} \dot{\theta}_m = \sum_n A_{m,n} \sin (\theta_n - \theta_m) 
                        + \frac{\widetilde{K}_s}{K} \sin(2 \theta_m).
\end{equation}
Here, $\theta $ is the phase of the $m $-th oscillator in the rotating frame
(with subtracted common linear in time contribution), $K$ is the coupling
parameter, matrix $\widehat{A} $ with matrix elements $A_{m,n} $ is the
network adjacency matrix, and $\widetilde{K}_s $ is the strength of the
phase
injection~\cite{adler_StudyLocking_1946,bhansaliGenAdler2009}.
By rescaling time, one can always choose $K=1 $. Since the analysis below
does not depend on the time scale, we simplify formulas by denoting
$K_s = \widetilde{K}_s/K $ and taking $K=1 $.

The equations of motion can be presented as induced  by the Lyapunov function 
\begin{equation}\label{eq:hqkm-def}
  H_K = \frac{1}{2} \sum_{m,n} A_{m,n} \cos(\theta_m - \theta_n)
  - \frac{K_s}{2} \sum_m \cos^2(\theta_m).
\end{equation}
Indeed, one can see that
$\dot{\theta}_m = - \partial H_K/\partial \theta_m $, implying that
$ dH_K/dt \leq 0 $. Thus, the system evolves in such a way that $H_K $
monotonously decreases, unless the system is in an equilibrium state, where
$\dot{\theta}_m = 0 $.

The extremal spectral properties of the Kuramoto model, the global minimum
of $H_K$ and the minimizing configurations, are tightly related to the
extremal properties of the Ising Hamiltonian $H_I$. This is the relation
between an integer program and its relaxations, as identified in the theory
of combinatorial optimization. To make this connection more
straightforward, we will consider it from the perspective of the max-cut
problem.

First, we observe that $H_K $ is equivalent to the Hamiltonian of the $XY $
model, which deals with unit vector distributions given by functions
$\mathbf{s} : \mathcal{V} \to \mathbb{S}^1 $, where $\mathbb{S}^1$ is the set of unit
2D vectors:
\begin{equation}\label{eq:xy-ham}
 H^{(XY)}(\mathbf{s}) = \frac{1}{2} \sum_{m.n} A_{m,n} \vec{s}_m \cdot \vec{s}_n - 
        \frac{1}{2} \sum_m {\left( \vec{l} \cdot \vec{s}_m \right)}^2,
\end{equation}
where $\vec{s}_m $ are unit vectors confined to a 2D plane, and $\vec{l} $
is the anisotropy axis. Defining the orientation of $\vec{s}_m $ in
terms of angle $\theta_m $ with respect to $\vec{l} $, so that we can write
$\vec{s}_m = {\left( \cos(\theta_m), \sin(\theta_m) \right)}^T $
in~\eqref{eq:xy-ham}, we obtain~\eqref{eq:qkm-eqm}.

Representing $H^{(K)} $ as a Hamiltonian of the XY model reveals the
connection between the dynamics of the Ising machine based on the
synchronizing oscillator network and the Ising model. The max-cut problem
can be presented as an integer
program~\cite{korte_CombinatorialOptimization_2018}
\begin{equation}\label{eq:maxcut-trace}
  \overline{C}_{\mathcal{G}}
       = \max_\Xi \left[ \frac{M}{2} 
         - \frac{1}{4}
         \mathrm{Tr}\left( \widehat{A} \widehat{\Xi} \right) \right],
\end{equation}
where matrix $\widehat{\Xi} $ is subject to constraints $\Xi_{i,i} = 1 $ and
$\mathrm{rank} (\widehat{\Xi}) = 1 $. It is easy to check that
$\widehat{\Xi}$ satisfies these constraints if and only if
$\Xi_{i,j} = \sigma_i \sigma_j $ with binary $\sigma_i $. Thus,~\eqref{eq:maxcut-trace} is
equivalent to the original formulation.

Let $C_{\mathcal{G}}^{(XY)}(\mathbf{s}) = M/2 - H^{(XY)}(\mathbf{s})/2$ be a formal
analog of cut for the XY model defined by using $H^{(XY)} $ instead of the
Ising Hamiltonian in Eq.~\eqref{eq:max-cut-ham} or, equivalently, by using
$\widehat{\Xi}^{(XY)}$ with $\Xi_{i,j}^{(XY)} = \vec{s}_i \cdot \vec{s}_j$ instead
of $\widehat{\Xi}$. One can see that the maximal value of
$C_{\mathcal{G}}^{(XY)}$ is found by solving Eq.~\eqref{eq:maxcut-trace} with the
relaxed rank constraint, $\mathrm{rank} (\widehat{\Xi}) = 2$, and requirement that
$\widehat{\Xi}^{(XY)}$ is positive semidefinite. Thus, the isotropic XY
model, and, hence, the isotropic Kuramoto model implement the rank-$2$
semidefinite programming (SDP) relaxation of the max-cut
problem~\cite{goemans_ImprovedApproximation_1995,
  alon_BipartiteSubgraphs_2000,
  deza_GeometryCuts_1997,Burer2002Rank-twoPrograms}.

\begin{figure}[!t]
  \centering
  \includegraphics[width=2.5in]{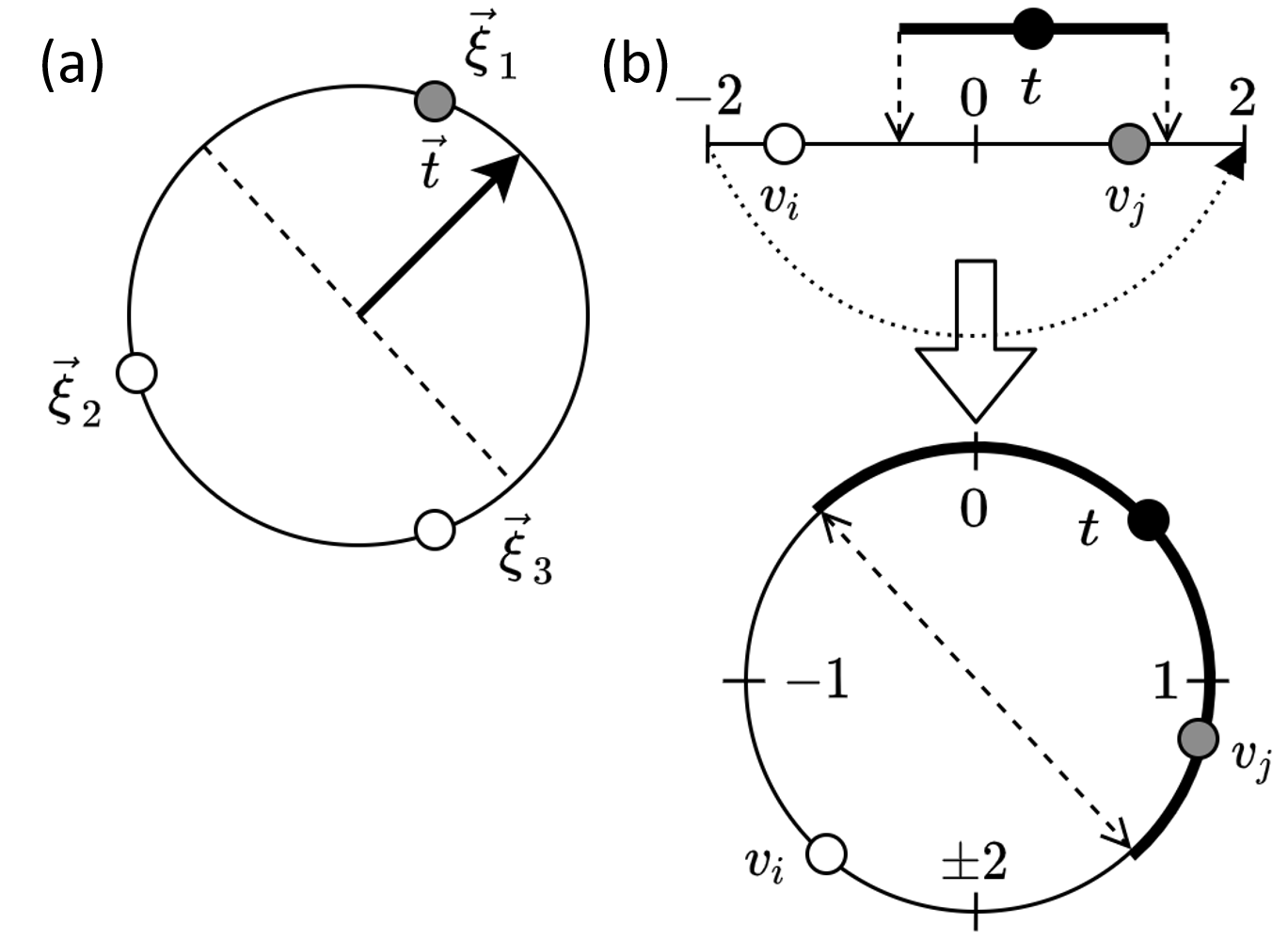}
  \caption{The rounding rules. (a) The rank-$2$ relaxation: vectors
    making positive dot product with $\vec{t}$ are rounded to $1$
    ($\vec{\xi}_1$, filled circles), and the rest are rounded to $-1$
    ($\vec{\xi}_2$ and $\vec{\xi}_3$, hollow circles). (b) The triangular
    model: the rounding center $t$ defines the center of the interval for
    mapping to $1$ ($v_j$ shaded circle), the rest of the variables are
    mapped to $-1$. }%
  \label{fig:rounding}
\end{figure}

One of the key elements of SDP relaxations is obtaining the binary state of
the Ising model, since finding the solution of the SDP
relaxation does not reduce the complexity of the max-cut
problem~\cite{laurent_positive_1995}. Let the relaxed problem be solved by
configuration $\mathbf{s}^{(0)}$ yielding
$\overline{\mathcal{C}}_{\mathcal{G}}^{(XY)} = C_{\mathcal{G}}^{(XY)}\left(\mathbf{s}^{(0)}\right)$.
Generally, vectors $\vec{s}^{(0)}_i $ are not oriented along the same line,
and, to reconstruct an Ising configuration, the XY configuration must be
rounded. This can be done by choosing unit vector $\vec{t} $ and mapping
vectors $\vec{s}^{(0)}_i $ to $\left\{ -1, 1 \right\} $ according to
$\sigma_i\left( \vec{t} \right) = \mathrm{sign}\left(\vec{s}^{(0)}_i \cdot \vec{t}
\right) $, as illustrated by Fig.~\ref{fig:rounding}(a).

Averaging obtained cut over orientations of $\vec{t}$
yields~\cite{goemans_ImprovedApproximation_1995}
\begin{equation}\label{eq:maxcut-aver}
 \left\langle  C_{\mathcal{G}}\left( \boldsymbol{\sigma}(\vec{t}) \right)  \right\rangle_{\vec{t}}
                \geq \alpha_{\text{G-W}} \overline{C}_{\mathcal{G}},
\end{equation}
where $\alpha_{\text{G-W}} = \min_{\theta > 0} \frac{\theta}{\pi \sin^2(\theta/2)} \approx 0.877\ldots$.

Inequality~\eqref{eq:maxcut-aver} guarantees that rounding the found
solution of the relaxed problem yields the cut at least within $13\% $ of
the maximal cut. This is the best, assuming that
$\mathrm{P} \ne \mathrm{NP}$, performance of an approximation to the max-cut
problem running in polynomial time.

Taking into account that
$\mathcal{C}_{\mathcal{G}}\left( \boldsymbol{\sigma}(\vec{t}) \right) $ cannot exceed the maximum
cut, we obtain an important inequality estimating the integrality gap, the
mismatch between the solutions of the relaxed and the original problems:
\begin{equation}\label{eq:mc-int-gap}
  \overline{C}_{\mathcal{G}} \geq 
  \alpha C_{\mathcal{G}}^{(XY)}\left( \mathbf{s}^{(0)} \right).
\end{equation}

Equations~\eqref{eq:maxcut-aver}--\eqref{eq:mc-int-gap} remain valid for
arbitrary non-negatively weighted adjacency matrices. The case, when both,
positive and negative, weights are allowed, is more complex. However, the
performance ratio and the integrality gap can be estimated in this case as
well~\cite{charikar_MaximizingQuadratic_2004,
  alon_ApproximatingCutnorm_2004, anjos_StrengthenedSemidefinite_2002}.

This consideration explains why dynamic Ising machines based on the
Kuramoto model can solve the max-cut problem and indicates that such machines
solvers can be quantitatively described by the target performance
guarantee. This is important because dynamical Ising machines that do not
converge to the Kuramoto model may leverage different computational
resources. Correctly identifying the origin of the computational power
allows recognizing the best area of applications, potential weaknesses and
challenges, and how to address them.

We conclude by noting that the gradient descent, generally, does not reach
the global minimum of $H^{(XY)}$~\cite{Burer2002Rank-twoPrograms,
erementchoukComputational2022}. At present, the common
approach to address this problem is to rerun the machine multiple times.
Alternative strategies for improving convergence can be explored but are
beyond the scope of the present paper.

\subsection{Almost-linear Ising machine}

The consideration above shows the natural connection between the SDP
relaxation and the dynamics of coupled phase oscillators with linear
coupling. Once the origin of the computational capabilities of oscillatory
IMs is established, one may reproduce it in a non-oscillatory
dynamical environment.

We consider an approach utilizing continuous unbounded dynamical variables,
which, for instance, can be represented by the electric charge. To make
such system behaving similarly to the XY model (and, hence, reproducing SDP
relaxations), one needs to emulate a cosine-like coupling between
individual variables. In the present paper, we avoid implementing such a
costly emulation by departing from the SDP approach in favor of a significantly
simplified model of coupling between the dynamical variables.

In the present paper, we explore a piece-wise linear approximation of the
coupling and the anisotropy functions in Eq.~\eqref{eq:qkm-eqm}. Using the
same conventions, $K_s = \widetilde{K}_s/K $ and $K = 1 $, the general
equations of motion are of the form
\begin{equation}\label{eq:IS-geneq}
 \dot{v}_i  = -\sum_j A_{i,j} \phi(v_i - v_j) + K_s \phi(2 v_i),
\end{equation}
where $v_i $ are dynamical variables with unrestricted values. The coupling
function, $\phi(v) $, is a piece-wise linear periodic function consistent with
a relaxation of the max-cut problem. We will slightly reduce the generality
of our consideration by requiring that $\phi(v) $ is an odd function.

To find $\phi(v) $, we consider the equations of motion as induced by the
Lyapunov function
\begin{equation}\label{eq:IS-gen-lyap}
  H^{(\Phi)}(\mathbf{v}) =
  \frac{1}{2} \sum_{i,j} A_{i,j} \Phi(v_i - v_j) - \frac{K_s}{2} \sum_i \Phi(2 v_i),
\end{equation}
where $\Phi(v) $ is an even periodic function related to the coupling function
by $ d\Phi(v)/dv = \phi(v) $, and satisfying $\Phi(0) = 1 $ and $\Phi(\pm 2) = -1 $.

Next, we define
\begin{equation}\label{eq:IS-gen-cut}
 C_{\mathcal{G}}^{(\Phi)} \left( \mathbf{v} \right) = \frac{M}{2} 
 - \frac{1}{4} \mathrm{Tr} \left( \widehat{A} \widehat{\Xi}^{(\Phi)} \right),
\end{equation}
where ${\Xi}^{(\Phi)}_{i,j} = \Phi(v_i - v_j) $. On feasible
($v_i = \pm 1 $) Ising configurations, $C_{\mathcal{G}}^{(\Phi)} $ coincides with the cut.
Thus, the problem
\begin{equation}\label{eq:IS-gen-maxcut}
 \overline{C}_{\mathcal{G}}^{(\Phi)} = \max_{ \mathbf{v} }  C_{\mathcal{G}}^{(\Phi)} \left( \mathbf{v} \right)
\end{equation}
is a proper relaxation of the max-cut problem and
$\overline{C}_{\mathcal{G}}^{(\Phi)} \geq \overline{C}_{\mathcal{G}} $.

\begin{figure}[!t]
        \centering
        \includegraphics[width=3in]{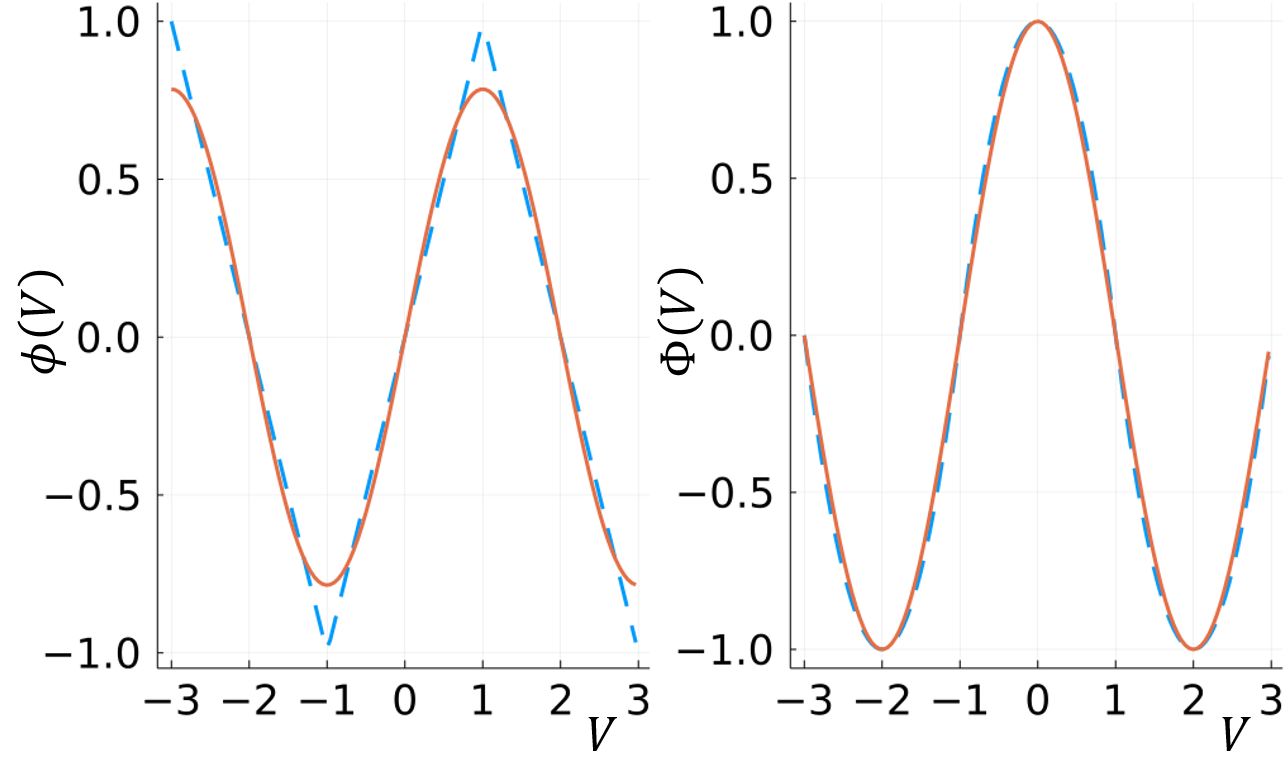}
        \caption{Comparison of the dynamic coupling function $\phi(v)$ and
        the Hamiltonian kernel $\Phi(v)$ for the rank-2 relaxation (solid
        line) and the triangular model (dashed line) }%
        \label{fig:st_comp}
\end{figure}

The simplest form of continuous with its derivative $\Phi(v) $ yielding a
piece-wise linear $\phi(v) $ is a function with the period $P = 4 $ and
defined within one period by
\begin{equation}\label{eq:IS-gen-Phi}
 \Phi(v) = 
        \begin{cases}
                1 - v^2, & -1 \leq v \leq 1, \\
                {(v - 2)}^2 - 1, & 1 \leq v \leq 3.
        \end{cases}
\end{equation}
This corresponds to 
\begin{equation}\label{eq:IS-gen-phi}
 \phi(v) = 2
        \begin{cases}
                - v, & -1 \leq v \leq 1, \\
                v - 2, & 1 \leq v \leq 3.
        \end{cases}
\end{equation}
We will call this relaxation the triangular model.
Figures~\ref{fig:rounding}(c,d) compare the coupling functions of the XY
and the triangular models showing the relationship between them.

The rounding procedure for the triangular model can be defined as follows.
Let $\mathbf{v}^{(0)}$ be a maximizing configuration so that
$\overline{C}_{\mathcal{G}}^{(\Phi)} = C_{\mathcal{G}}^{(\Phi)} \left( \mathbf{v}^{(0)} \right)$.
Because of the periodicity of $\Phi(v) $, we can take $v_i^{(0)}$ modulo
$P $ and restrict them to the single period, say, $\left[ -2, 2 \right] $,
with identified ends. Such mapping of $v_i^{(0)}$ makes the rounding
procedure for the triangular model essentially the same as for the XY model
as illustrated by Fig.~\ref{fig:rounding}(a). For the chosen rounding
center $t$, those $v_i $ that fall inside the interval
$\left[ t - P/2, t+P/2 \right) $ are mapped to $+1 $, while the rest are
mapped to $-1 $. We denote the rounding mapping for the given rounding
center $t$ by
$\boldsymbol{\sigma} = \widehat{R}\left[ \mathbf{v}^{(0)} ; t \right]$.

Moreover, the performance and the integrality gap can also be
estimated using the same argument as for the SDP relaxations. Let
$\boldsymbol{\sigma}(t)$ is the Ising configuration obtained by mapping
$\mathbf{v}^{(0)}$ to $\left\{ -1, 1 \right\}^N$ with the given the
rounding center $t $. Averaging the resultant cut over the random choices
of $t $, we find
\begin{equation}\label{eq:IS-gen-av}
 \left\langle C_{\mathcal{G}} \left( \boldsymbol{\sigma}(t) \right) \right\rangle_t
                = \sum_{i,j} A_{i,j} \mathcal{P}\left(v_i^{(0)} , v_j^{(0)}\right),
\end{equation}
where $\mathcal{P}\left(v_i^{(0)} , v_j^{(0)}\right) $ is the probability that
$v_i^{(0)} $ and $v_j^{(0)} $ fall into different intervals and, hence,
edge $(i,j) $ contributes to the cut. Because of the rotational symmetry of
the rounding, this probability may depend only on the mutual arrangement of
$v_i^{(0)} $ and $v_j^{(0)} $. Thus, we have the usual geometric
probability situation and can write
\begin{equation}\label{eq:IS-gen-pij}
  \mathcal{P}\left(v_i^{(0)} , v_j^{(0)}\right) =
           \frac{2}{P} \left\vert v_i^{(0)} - v_j^{(0)} \right\vert_P,
\end{equation}
where $\left\vert v_i^{(0)} - v_j^{(0)} \right\vert_P$ is the distance between
$v_i^{(0)} $ and $v_j^{(0)}$ along the circumference in
Fig.~\ref{fig:rounding}(a) (the length of the shortest path).

Thus, we have
\begin{equation} \label{eq:IS-gen-cutaver}
\begin{split}
    \left\langle C_{\mathcal{G}} \left( \boldsymbol{\sigma}(t) \right) \right\rangle_t
            &    = \sum_{i,j} A_{i,j} \frac{2}{P} \vert v_i^{(0)} - v_j^{(0)}\vert_P \\
            & \geq \alpha_T \sum_{i,j} A_{i,j} \frac{1}{2}
                       \left[ 1 - \Phi \left(v_i^{(0)} - v_j^{(0)} \right) \right] \\
            &\geq \alpha_T \overline{C}_{\mathcal{G}}^{(\Phi)} \geq \alpha_T \overline{C}_{\mathcal{G}},
\end{split}
\end{equation}
where
\begin{equation} \label{eq:IS-gen-alphaT}
    \alpha_T = \frac{4}{P} \min_{\vert v\vert \leq 2} \frac{\vert v\vert}{1 - \Phi(v)} \approx \frac{4}{P} 0.85\ldots .
\end{equation}
This estimates the integrality gap of the triangular model and its enabling
max-cut performance.

Above, we kept explicitly period $P$ of the coupling function to show that
$P = 4$ is, in a sense, optimal. Such an optimization, of course, is
trivial from the perspective of various nonlinear relaxations. Even when
restricted to translationally invariant $\Phi(v)$ yielding piece-wise linear
$\phi(v)$, the family of these relaxations is very rich from the perspective
of enabled dynamics and performance consequences. The respective family of
dynamical Ising machines based on different coupling functions
$\Phi(v, v')$ is yet to be explored. In the present paper, we limit ourselves
to the triangular model as a particular representative of the extended
family.


\section{Software simulation of the almost-linear Ising machine}
\label{sec:software-simulation}

In this section, we present the software simulation of the Ising machine
based on the triangular model. It should be noted that there are various
ways of improving the practical performance of the Ising machine. For
example, the adaptive dynamics can be used for speeding up the convergence
of the machine to a steady state, restarting the machine from a slight
perturbation of the previously found the best configuration and special
schedules for time varying anisotropy constant, $K_s$, can improve the
search for more optimal solutions, and so on. A detailed analysis of such
techniques, however, is beyond the scope of the present paper, and we focus
on studying the core performance of the Ising machine based on the
triangular model.

\begin{figure}[!t]
  \centering
  \includegraphics[width=1.25in]{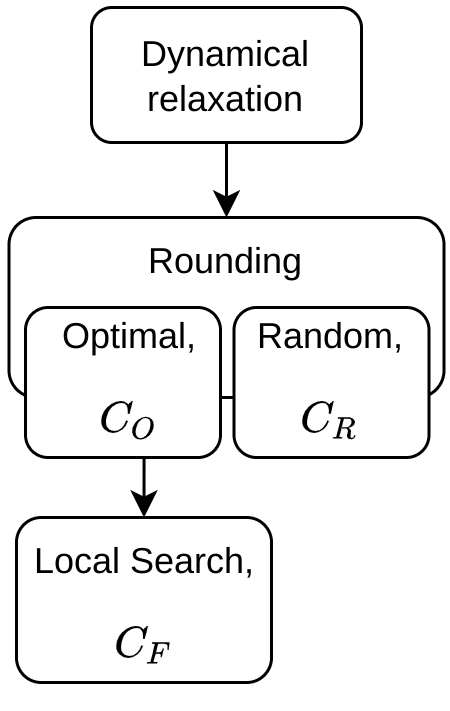}
  \caption{Three setups testing the basic performance of the Ising
    machine based on the triangular model}
  \label{fig:simulation}
\end{figure}

The general structure of simulations is shown in Fig.~\ref{fig:simulation}.
Testing consists of three stages. During the first stage, the Ising machine
evolves freely according to the equations of motion solved using the Euler
approximation. During the second stage, the machine state established after
the \emph{fixed} number of time steps is rounded to construct a binary
state of the Ising model. As has been discussed above, rounding is one of
the key elements of finding the maximal cut using the SDP relaxation.
Therefore, we compare two approaches to rounding based on mapping
$\widehat{R}$ defined above.
 
One approach leverages the observation that averaging cut over randomly
chosen rounding centers is sufficient for getting the best performance
assuming that $\text{P} \ne \text{NP}$~\cite{khot_optimal_2004}. Within this
approach, the rounding center is sampled $n_R$ times from the uniform
distribution over the interval $\left[ -1, 1 \right]$. The 
configuration is rounded according to the chosen values of the rounding
center and the best value of cut out of $n_R$ is kept.

\begin{algorithmic}[0]
  \Procedure{Random-Rounding}{input $\sset{G}$, $\mathbf{v}$, $n_R$; output
    $C_R$, $\boldsymbol{\sigma}_R$}
  \State $C_R \gets -1$
  \State Let $\left\{ t_1, \ldots, t_{n_R} \right\}$ be a sample of the uniform
  distribution on $\left[ -1, 1 \right]$
  \ForAll{$t_i \in \left\{ t_1, \dots, t_{n_R} \right\}$}
     \State $\boldsymbol{\sigma}(t_i) \gets \widehat{R}[\mathbf{v}; t_i]$
     \State $C(t_i) \gets C_{\sset{G}}\left( \boldsymbol{\sigma}(t_i) \right)$
     \If{$C(t_i) > C_R$}
     \State $C_R \gets C(t_i)$
     \State $\boldsymbol{\sigma}_R \gets \boldsymbol{\sigma}(t_i)$
     \EndIf
  \EndFor
  \EndProcedure
\end{algorithmic}

\begin{figure}[ht]
  \centering
  \includegraphics[width=3.5in]{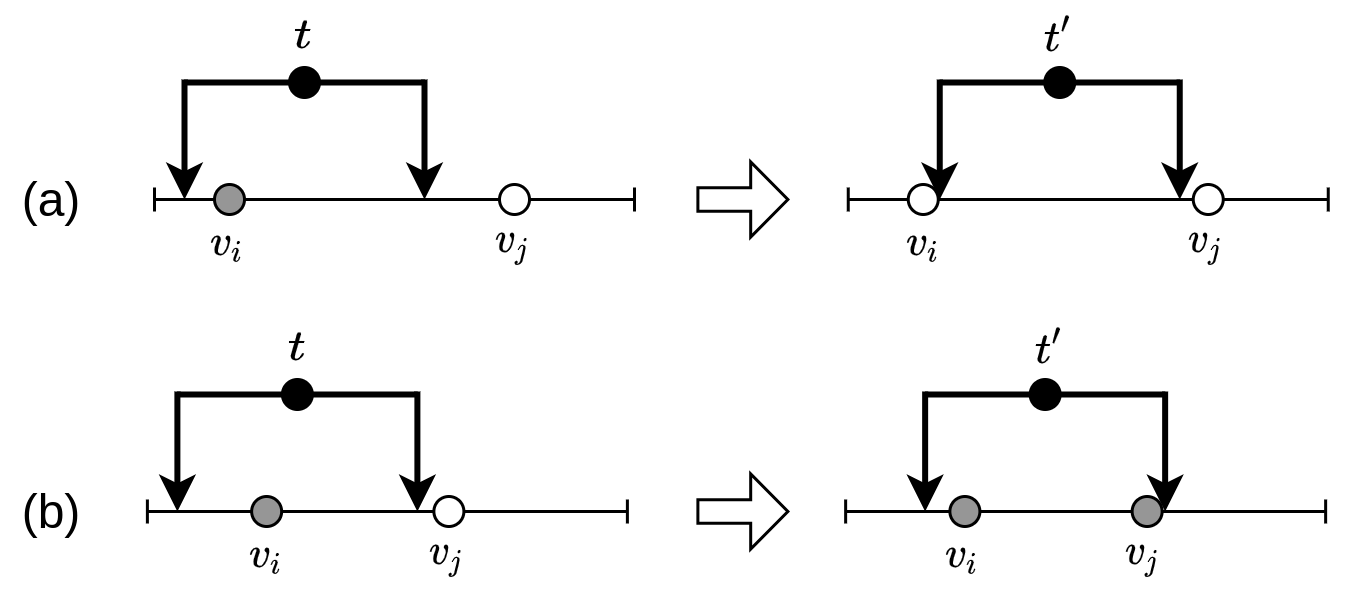}
  \caption{\label{fig:opt-rounding} The rounding center can be shifted in
  a way ensuring that only one spin in the rounded configuration is reverted.
This happens when either (a) the left boundary encounters the node
previously mapped to $+1$, or (b) the right boundary reaches the node
previously mapped to $-1$.}
\end{figure}

The second approach evaluates all non-equivalent binary states that can be
obtained by varying the rounding center. This can be achieved by monotonously
traversing the interval $[-1, 1)$. Ordering coinciding $v^{(0)}_i$
according to the nodes enumeration, one can ensure that the recovered
partitions change by a single node as illustrated by
Fig.~\ref{fig:opt-rounding}.

Let the index of the reversed spin be $p$, so that its state changes from
$-\sigma_p$ to $\sigma_p$ after displacing $t$. The change of cut is then
\begin{equation}\label{eq:delta-c}
  \Delta C_p = C(\left\{ \sigma_p \right\} ) - C(\left\{ -\sigma_p \right\} )
   = F_p(\boldsymbol{\sigma}),
\end{equation}
where
\begin{equation}\label{eq:fp-imbalance}
  F_p(\boldsymbol{\sigma}) = - \sum_{n} A_{p,n} \sigma_p \sigma_n
\end{equation}
is the imbalance between cut and uncut edges incident to node $p$. Thus,
when $t$ reaches $t'$, the total cut variation is
\begin{equation}\label{eq:delta-c-total}
  \Delta C(t') =\sum_{k = 1}^{n(t')} \Delta C_{p_k},
\end{equation}
where $p_1, \ldots, p_{n(t')}$ is the sequence of spins inverted while $t$ was
traversing from $-1$ to $t'$. The optimal rounding is determined by the
position of the rounding center $t_O$, at which $\Delta C(t)$ takes the
maximal value for $-1 \leq t < 1$.

\begin{algorithmic}[0]
  \Procedure{Optimal-Rounding}{input $\sset{G}$, $\mathbf{v}$; output
    $C_O$, $\boldsymbol{\sigma}_O$}
  \State $t \gets \max \left( -1, \min_i (v_i - 1) \right)$
  \State $t_O \gets t$, $\Delta C_{\text{max}} \gets -1$, $\Delta C \gets 0$
  \State $\boldsymbol{\sigma} \gets \widehat{R}[\mathbf{v}; t_O]$
  \While{$t < 1$}
      \State $a \gets \min \left(-1, \arg \min \left(v_i : t-1 \leq v_i < t+ 1 \right)\right)$
      \State $b \gets \min \left(1, \arg \min \left(v_i : t + 1 \leq v_i \right) \right) $
      \If{$v_a + 1 \leq v_b - 1$}
          \State $p \gets a$
      \Else
          \State $p \gets b$
      \EndIf
      \State $t \gets v_p$, $\sigma_p \gets -\sigma_p$
      \State $\Delta C \gets \Delta C + F_p(\boldsymbol{\sigma})$
      \If{$\Delta C > \Delta C_{\text{max}}$}
          \State $\Delta C_{\text{max}} \gets \Delta C$
          \State $t_O \gets t$
      \EndIf
  \EndWhile
  \State $\boldsymbol{\sigma}_O \gets \widehat{R}[\mathbf{v}; t_O]$
  \State $C_O \gets  C_{\sset{G}}\left( \boldsymbol{\sigma}_O\right)$
  \EndProcedure
\end{algorithmic}

\begin{figure}
    \centering
    \includegraphics[width=3in]{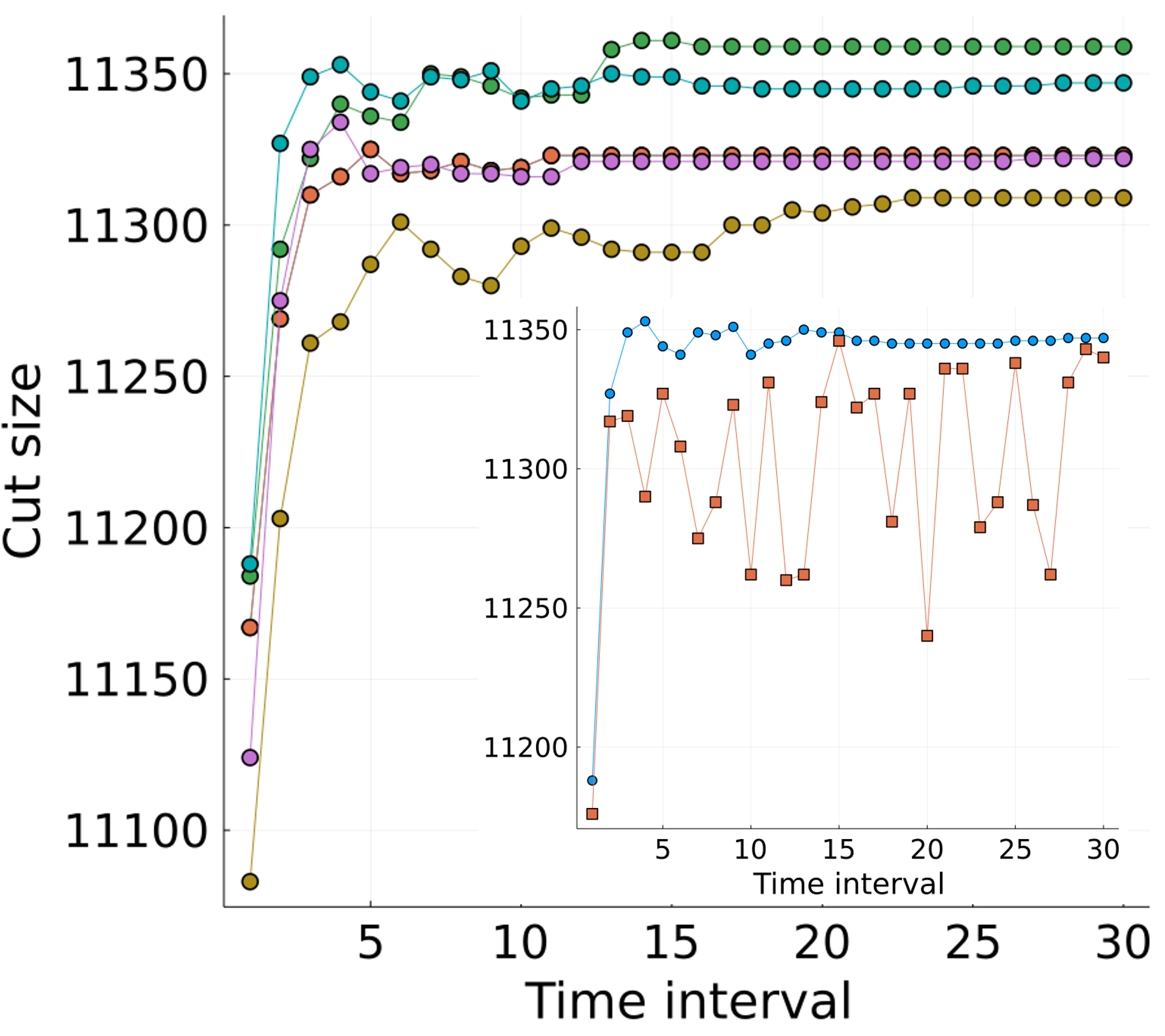}
    \caption{The time evolution of the cut of graph $\mathrm{G}1$.
      The main panel shows the results of the evolution starting from five
      random initial configurations. The inset compares the optimal rounding
      (marked by circles) with the random rounding (marked by squares). }%
    \label{fig:tri-cut-evolution}
\end{figure}

Figure~\ref{fig:tri-cut-evolution} compares optimal and random roundings
for the example of graph $\mathrm{G}1$ from \texttt{Gset}~\cite{gset}.
This figure shows the triangular model evolving freely with snapshots of
its configuration taken after $600K/N$ long time intervals. The time
evolution has the characteristic form: fast initial growth followed by very
slow dynamics signifying that the machine is near a steady state. The
diminishing returns in increasing the machine running time and the
significant variation of the cuts between individual runs suggest that
efficient dynamics control are advantageous to the machine practical
performance as has been mentioned above.

In the simulation outlined in Fig.~\ref{fig:simulation}, the Ising machine
runs freely for $250$ time steps, each $140K/N$ long. The binary state
obtained after optimal rounding are processed further using the local
search. First, it is ensured that all nodes satisfy the node majority rule
(NMR), that is $F_m \geq 0$ for all $m \in \sset{V}$.  Then, it is ensured
that cut edges obey the edge majority rule (EMR), which emerges while
applying NMR to the pairs of nodes incident to a cut edge. Let edge $(i,j)$
be cut. Then, the number of cut edges adjacent to $(i,j)$ must exceed the
number of uncut edges at least by $2$. Otherwise, reverting $\sigma_i$ and $\sigma_j$
will increase cut. We denote the finally obtained cut by $C_F$. The whole
procedure is repeated $100$ times and the best results are kept.

The results of the simulation on $\left\{ 0, 1 \right\}$-weighted graphs from
\texttt{Gset}~\cite{gset} are presented in
Table~\ref{tab:models-benchmarks}, where they are compared with solutions
of the max-cut problem found by optimizer
\texttt{Circut}~\cite{Burer2002Rank-twoPrograms} implementing rank-$2$ SDP
followed by the local search (the same as described above). While not all
\texttt{Circut} numbers are the best among known for \texttt{GSet} (see,
for example,~\cite{benlic_BreakoutLocal_2013}), the deviations are
sufficiently small.

\begin{table*}[t]
\caption{The performance of the Ising machine based on triangular model compared
  to \texttt{Circut}~\cite{Burer2002Rank-twoPrograms}}%
\label{tab:models-benchmarks}
\centering
\begin{tabular}{l|l|l|l|l}
\hline
\multirow{2}{*}{Test graph 
        ($N$ : $M$)} & \multicolumn{3}{c|}{Triangular model machine} & \multirow{2}{*}{\texttt{Circut}} \\ \cline{2-4}
                     & Random, $C_R$      & Optimal, $C_O$      & Processed, $C_F$  & \\ \hline
G1 (800 : 19176)     & 10052     & 10113       & 11524       & 11624                   \\
G2 (800 : 19176)     & 9984      & 9993       & 11534       & 11620                   \\
G3 (800 : 19176)     & 10010     & 10034       & 11447       & 11622                   \\
G4 (800 : 19176)     & 10308     & 10379       & 11582       & 11646                   \\
G5 (800 : 19176)     & 10088     & 10146       & 11522       & 11631                   \\
G22 (2000 : 19990)   & 13066     & 13092       & 13249       & 13353                   \\
G23 (2000 : 19990)   & 13068     & 13084       & 13202       & 13332                   \\
G24 (2000 : 19990)   & 13038     & 13061       & 13207       & 13324                   \\
G25 (2000 : 19990)   & 13042     & 13046       & 13239       & 13329                   \\
G26 (2000 : 19990)   & 13044     & 13054       & 13225       & 13321                   \\
G43 (1000 : 9990)    & 6334      & 6348        & 6604        & 6659                    \\
G44 (1000 : 9990)    & 6312      & 6321        & 6591        & 6648                    \\
G45 (1000 : 9990)    & 6343      & 6347        & 6594        & 6653                    \\
G46 (1000 : 9990)    & 6341      & 6358        & 6585        & 6645                    \\
G47 (1000 : 9990)    & 6343      & 6391        & 6573        & 6656                    \\
G48 (3000 : 6000)    & 5722      & 5728        & 5746        & 6000                    \\
G49 (3000 : 6000)    & 5750      & 5752        & 5774        & 6000                    \\
G50 (3000 : 6000)    & 5690      & 5694        & 5736        & 5880                    \\
G51 (1000 : 5909)    & 3640      & 3659        & 3786        & 3846                    \\
G52 (1000 : 5916)    & 3650      & 3666        & 3792       & 3847                    \\
G53 (1000 : 5914)    & 3660      & 3672        & 3793        & 3846                    \\
G54 (1000 : 5916)    & 3773      & 3667        & 3788        & 3850                    \\ \hline
\end{tabular}
\end{table*}

\begin{figure*}[!t]
  \centering
  \includegraphics[width=0.9\textwidth]{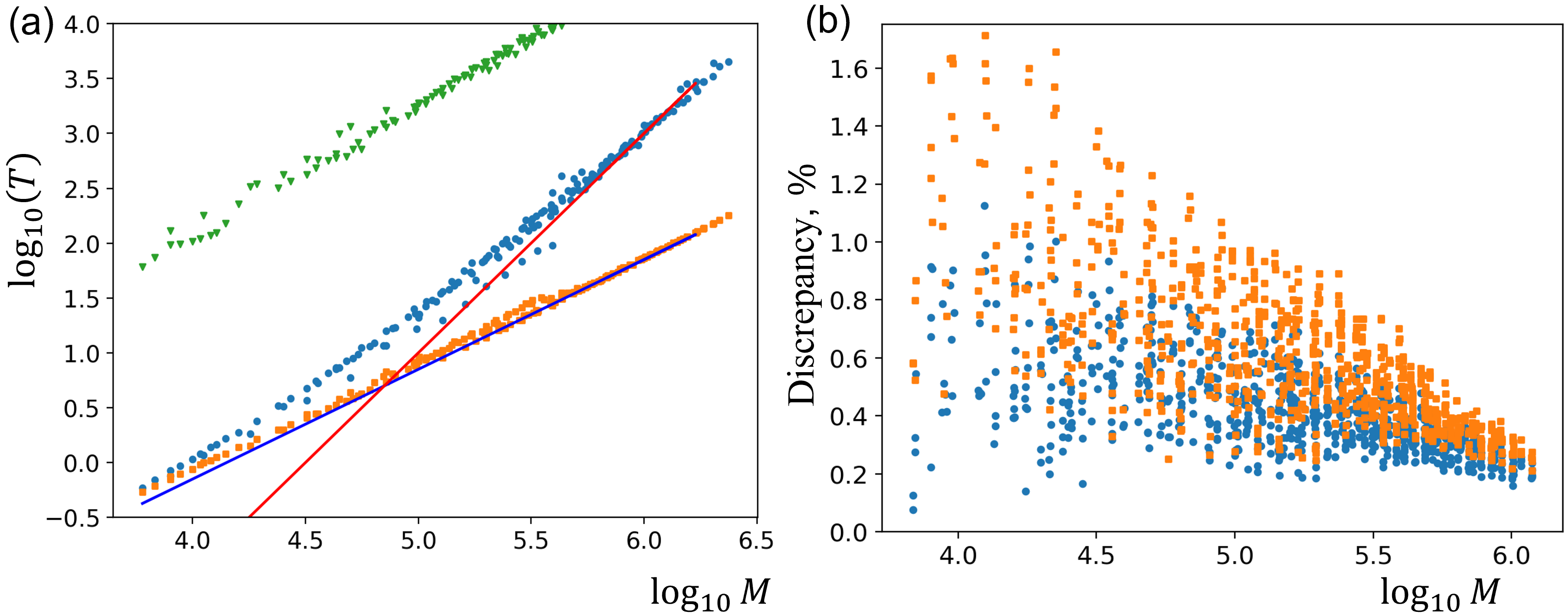}
  \caption{(a) The variation of the running time (the wall-time measured in
    seconds) with increasing of the size for an ensemble of random Erd\H{o}s-R\'{e}nyi
graphs $\sset{G}_{N,p}$: (circles) full, NMR and EMR, post-processing,
(squares) reduced, NMR only, post-processing, (triangles) \texttt{Circut}'s
results. The straight lines show scaling $T \propto M^2$ and $T \propto M$. (b) The
discrepancy between the results using the almost-linear Ising machine with
(circles) full and (squares) reduced post-processing.}
  \label{fig:tri-scaling-disc}
\end{figure*}

To estimate how the running time of the machine scales with the size of the
problem and how the processing techniques affect the practical machine
performance, we apply the machine to a series of random Erd\H{o}s-R\'{e}nyi
graphs $\sset{G}_{N,p}$, where $N$ is the number of graph nodes, and $p$ is
the edge presence probability. The ensemble was obtained by sampling five
graph for each $N$ in $\left( 200, 400, 600, \ldots, 4000 \right)$ and $p$ in
$\left( 0.05, 0.1, \ldots, 0.35 \right) \cup \left( 0.12, 0.17, \ldots, 0.32 \right)$.
In each simulation, the machine ran freely from a weak perturbation of the
best configuration found so far (for the first run of each graph, the best
configuration was taken with all spins up). The machine ran $50$ time-steps,
each $50\ K/N$ long. The resultant state was optimally rounded and then
further post-processed. The found cut is compared to the best found. For
each graph, the procedure was repeated $30$ times. The results of the
simulation are shown in Fig.~\ref{fig:tri-scaling-disc} depicting the
running time and the deviation of the obtained cut values from the
\texttt{Circut}'s results versus the number of edges in the graphs.

The dynamical part of looking for the maximal cut requires the fixed number
of steps with the number of arithmetic operations determined by the number
of edges and, therefore, scales as $\mathcal{O}(M)$. The optimal rounding procedure
evaluates the variation of cut while reverting $N-1$ spins and thus also
scales as $\mathcal{O}(M)$.

The main contribution of the post-processing procedures to the running time
scaling is due to enforcing the EMR, which evaluates the vicinities of cut
edges, whose number scales as $M$, and, consequently, has
\emph{the-worst-case-scenario} scaling $\mathcal{O}(M^{5/2})$. To estimate
the effect of enforcing the EMR for Erd\H{o}s-R\'{e}nyi graphs, we compare in
Fig.~\ref{fig:tri-scaling-disc} implementations including both
post-processing procedures and enforcing EMR only. The numerical data shows
the transition to the faster running time growth (from $\mathcal{O}(M)$ to
$\mathcal{O}(M^2)$). It must be emphasized that for graphs of moderate size
($M > 10^6$) abandoning the last step in post-processing speeds up the
machine by more than the order of magnitude while leading to a fraction of
percent loss of accuracy.

\section{CMOS implementation of the almost-linear Ising machine}

The almost-linear character of the dynamics of the Ising machine
significantly simplifies a circuit implementation, which can be based on
the Euler approximation of the almost-linear Ising machine described by
Eq.~\eqref{eq:IS-geneq}:
\begin{equation}
  \Delta v_i\left[n+1\right]=
  -K \sum_{j=1}^N w_{ij} \phi\left(v_i[n]-v_j[n]\right)
  + K_s \phi\left(2v_i[n]\right).
    \label{eq_discr_state}
\end{equation}
Here, $v_i$ and $v_j$ are dynamical variables, $w_{ij}$ is the
weight parameter, $n$ is the time-step number, and $\phi(v)$ is the triangular
coupling function defined in Eq.~\eqref{eq:IS-gen-phi}.

\subsection{Architectural overview}

\begin{figure*}
    \centering
    \includegraphics[width=0.95\textwidth]{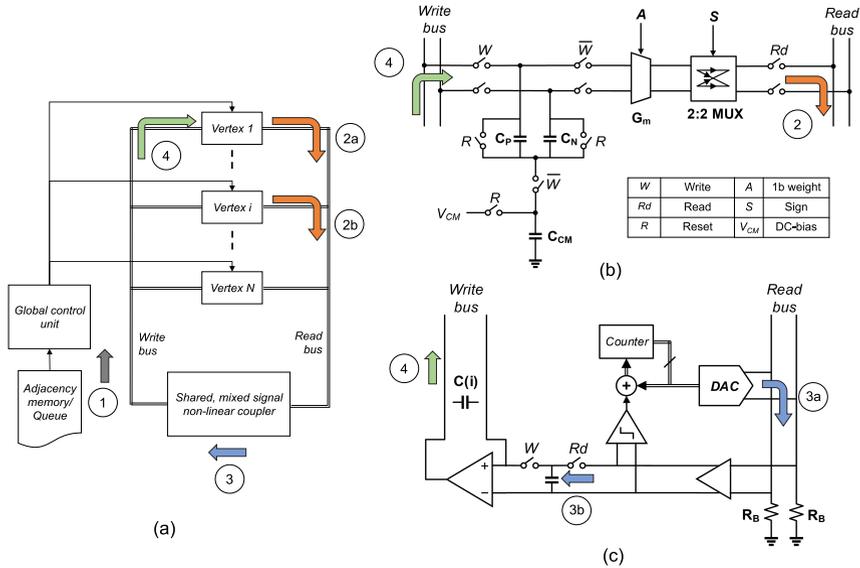}
    \caption{The proof-of-concept implementation of the almost linear Ising
      machine. (a) Top-level
      block-diagram of the solver and enumerated signal-flows. (b) Vertex
      schematic, with the signals of (a) identified. Circuit elements are
      labeled in bold and binary control-signals in italics. The table
      lists the signals' names. Any control signal's ON state closes the
      corresponding switch. (c) Non-linear coupler's block-diagram, with
      the signal-flows of (a) identified. Step 3 is split into two
      sub-steps, 3a and 3b. 3a is the application of current components and
      3b is the buffering of bus-voltage into the input capacitor.}%
    \label{sysdesc}
\end{figure*}

The essential components of the circuit implementation of the almost-linear
Ising machine are the vertex elements, a shared (periodic) coupler and the
adjacency memory. Their interconnection is shown in Fig.~\ref{sysdesc}(a).

Inside each vertex, a capacitor is employed as the analog spin memory
functioning as an analog accumulator. We use a differential current-mode
read of the vertices because the majority of operations in
Eq.~\eqref{eq_discr_state} can be broken down into several summations and
sign-inversions of analog signals. The first key operation in
Eq.~\eqref{eq_discr_state} is the pairwise subtraction of states. Currents
from a pair of vertices are easily subtract if the sinking node is held at
a nearly fixed voltage. The second key operation is the piece-wise linear
triangular coupling function ($\phi$). A current-mode read enables a
straightforward realization of $\phi$, as it may be described as a sum of
uniformly spaced currents followed by sign-inversions.

A current-bus runs next to all vertices, (i) to carry the current
originating from the vertices towards the coupler, and (ii) to carry the
state increments ($\Delta v_{ij}$) from the coupler to the vertices. As shown in
Fig.~\ref{sysdesc}(a), the bus is split into a write-bus and read-bus for
simplifying the decoding logic in the spin cell. To realize the triangular
coupling ($\phi$), the coupler senses the net current on the bus and alters it
by generating and superposing its own current. Additional circuit
components, discussed later, translate the net bus-current to state
increments. The adjacency memory stores weights from the weighted adjacency
matrix ($w_{ij}$) of the graph.

The key operations, depicted in Fig.~\ref{sysdesc}(a--c), are:
\begin{enumerate}[(i)]
\item reading the adjacency memory to get $\{i,j\}$ pair for each
  $w_{ij}=1$,
\item enabling the reading of vertices $i$ and $j$, by enabling the
  corresponding trans-conductance (Gm) cells,
\item application of the coupling current and reading the steady-state
  current on the bus,
\item updating the state capacitor's charge of vertex $i$.
\end{enumerate}

The circuit-level details of each of the components are discussed next.

\subsection{Design methodology and details}

\subsubsection{Vertex}

\begin{figure}
    \centering
    \includegraphics[scale=0.50]{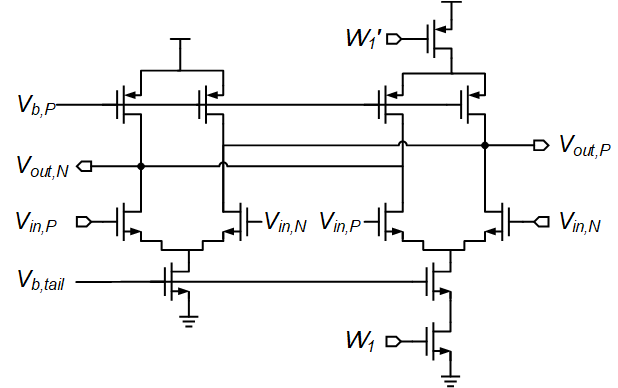}
    \caption{$1$-bit Gm-cell's transistor-level schematic. $V_{in,P}$ and
      $V_{in,N}$ are the differential inputs, $V_{out,P}$ and $V_{out,N}$
      the differential outputs, $W_1$ the digital input, $V_{b,P}$ and
      $V_{b,tail}$ the global bias-voltages. The left branch is active
      regardless of digital input. The second branch is active for
      $W_1=1$.}%
    \label{gm_schem}
    \end{figure}

The schematic along with the operational signals of the vertex --- read,
write, reset, labelled as R, W, and Rd are shown in Fig.~\ref{sysdesc}(b), next
to the corresponding switches. The vertex comprises:
\begin{enumerate}[(i)]
\item \emph{State capacitors}: Three capacitors, arranged in a T, are used
  to store the analog states. $C_P$ and $C_N$ store charge/voltage in a
  differential form. $C_{CM}$ stores the common-mode voltage that provides
  a steady bias for the Gm-cell. Reset switches reset $C_P$ and $C_N$ to
  have zero charge and $C_{CM}$ to $V_{CM}$. A switch blocks the
  charge/discharge of $C_{CM}$ during the update-phase and is enabled only
  during the read-phase.
  \item \emph{1-bit differential Gm-cell}: The cell reads the state voltage
    and supplies a proportional current onto the bus. To conserve area, we
    use a basic differential-pair with a tail source (Fig.~\ref{gm_schem}).
    The input transistor pair is biased by the common-mode capacitor. The
    tail-source is biased externally, with a global biasing voltage. Two
    identical differential pairs are used, where the second pair, when
    enabled, allows pre-multiplication with a factor two of the anisotropy
    term of Eq.~\eqref{eq_discr_state}.
  \item \emph{Sign-inverter}: The opposite signs of $v_i$ and $v_j$ in
    Eq.~\eqref{eq_discr_state} necessitates means to invert the
    sign/direction of vertex current. For inverting the currents, a 2-way
    in and 2-way out MUX was used which swaps the current into the bus.
  \item \emph{Vertex read-enable switch}: This switch enables/disables the
    vertex current onto the bus.
\end{enumerate}    

Assuming DRAM capacitor size of 10$\lambda^2$ requires refresh time of 1
$\mu$s, where $\lambda$ is the minimum feature size, we use capacitors with size at
least $10^3\times$ bigger than DRAM size, so that it works without refreshing as
it operates for the worst-case time-scale of $1$ ms. For a commercial $130$
nm technology, this is equivalent to a capacitor of about $10$ fF and an
area of at least $10^4 \lambda^2$.

\subsubsection{Spin-coupler}

It consists of three components: digital counter, comparator, and
sign-inverter. The counter, depending on the comparator output, changes its
state to execute the following operation:
\begin{equation}
    \min_{m}\left\vert I_i - I_j - mI_0 \right\vert,
\end{equation}
where, $m$ is an integer denoting the counter's state, $I_i$ and $I_j$
denote the currents from the vertices and $I_0$ denotes the smallest
positive zero of the coupling function. Instead of linear search, faster
search methods may be used. As shown in the Fig.~\ref{sysdesc}(c), counter
value is converted to an analog current and added to the net
branch-current. An additional current inverter is enabled when $m$ is an
even number to invert the slope of the linear segments in the coupling
function.

The number of segments in the coupling function determines the number of
bits required by the counter and by the DAC. It can be shown that if the
function is to be composed of $n$ periods, then $\lceil{\log_2{4n}}\rceil$ bits are
required. If charge increment at each time-step is sufficiently small,
limited number of periods (or, peaks) in the non-linear coupling function
are required. For our demonstration, we used 3 periods, or a 4-bit DAC.

\subsubsection{Switched capacitor based accumulator}

To implement the accumulation of Eq.~\eqref{eq_discr_state}, we use a
simplified switched capacitor setup~\cite{Razavi2002DesignCircuits} --- an
operational amplifier with single-ended output, a buffer/input capacitor
and the state-capacitor as the accumulator (Fig.~\ref{sysdesc}(c)). At each
sampling-instant, an input capacitor stores the amplified output of the
current-bus (step 3b in Fig.~\ref{sysdesc}(c)). This is followed by a
writing-instant, wherein the same charge is pushed up into the
state-capacitor (step 4 in Fig.~\ref{sysdesc}(c)).

To implement different charge increments for the coupling and anisotropy
term ($K$ and $K_s$ terms in Eq.~\eqref{eq_discr_state}, respectively), two
different capacitors must serve at the input of the op-amp. For this
reason, we use a programmable capacitor combination (Fig.~\ref{gm_schem}).
Since $K_s$ is typically smaller than $K$, for anisotropy increments, a
smaller input capacitor in series with the branch is activated so that a
relatively small charge is pushed into the vertex. Similarly, the overall
scale of the increments may be decreased for larger graphs by activating
even smaller capacitors in series with the branch.

\begin{figure}
    \centering
    \includegraphics[scale=0.45]{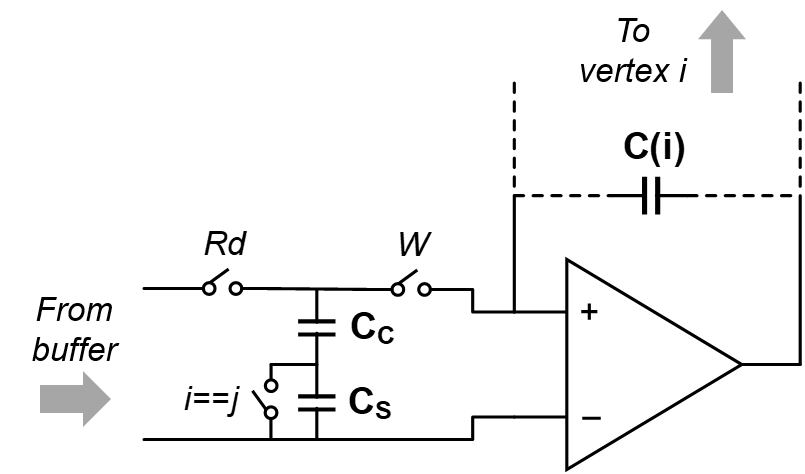}
    \caption{Switched capacitor arrangement for realizing accumulation.
      `Rd' pulse mirrors the bus voltage as a charge. `W' pulse pushes the
      entire charge into the state-capacitor. For anisotropic increments
      ($i=j$), the input capacitance is adjusted appropriately.}%
    \label{sw_cap_KsK}
\end{figure}

\subsubsection{State-initializer}

In the vertex element presented above, the initial charge on the state
capacitor is set to zero which, according to Eq.~\ref{eq_discr_state}, may
prevent the system from ever evolving. For this reason and to facilitate a
better exploration of the energy landscape, the state-voltages must be set
to random values. We use a 4-bit pseudo-random bit sequence generator
(PRBSG) to generate a random bit sequence. A shared DAC converts PRBSG
output to the state-capacitor's voltage. Since the output sequence of PRBSG
consists of equi-probable 1 and 0, the initial charge on the
state-capacitor is uniformly distributed. It suffices for the initial
capacitor voltage to be within the first linear segment of the
signed-modulo coupling function. Therefore, all of the bits input to
DAC are of lower significance than the bits used in the periodic coupler.

\begin{figure}
\centering
\includegraphics[scale=0.43]{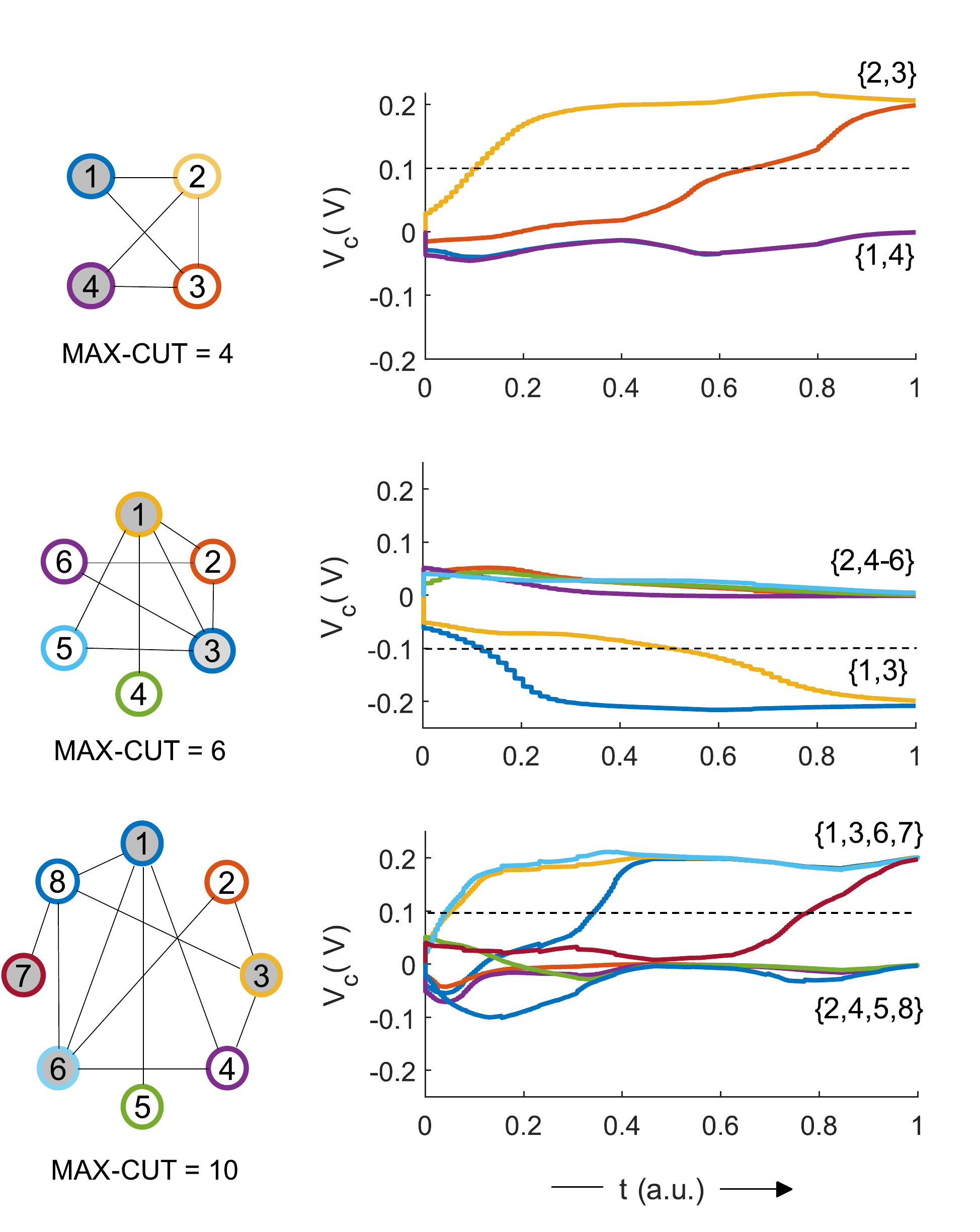}
\caption{Max-cut results obtained for the three
  graphs shown on the left. The time variation of state
  variables corresponding to each vertex is shown on the right by curves of
  matching colors. The shaded and hollow vertices belong to different sets
  in the max-cut partition. The dashed lines show the threshold capacitor
  voltage separating vertices between the partitions.}%
\label{fig:spice_results}
\end{figure}

\subsection{Software-equivalence test}

We used the IM for solving max-cut problem on randomly connected graphs.
Due to long simulation times, we simulated up to 8 nodes, each with a
random initial state. Figure~\ref{fig:spice_results} plots the evolution of
analog spins ($V_C$) corresponding to each vertex. On reaching the zeros of
the coupling function (200 mV), state-voltages did not change
significantly. The partition corresponding to the maximal-cut was found
using thresholds of $\pm100$ mV. With $K_s/K$ in Eq.~\eqref{eq_discr_state}
set to 1, one run was enough for finding the maximal cut for the selected
graphs. Correct determination of the max-cut values for the graphs verifies
the software-equivalence for the proposed hardware.

The proposed design can be further refined to
facilitate large node integration. For instance, the current bus can be
segmented to improve coupling between on-chip distant nodes.
Another refinement could involve the use of alternative methodologies to
activate vertices. The current design uses de-multiplexers to
enable/disable vertex reads, and, as a result, only one vertex can be read
at a time. A serial communication scheme may be employed to enable reading
multiple vertices at a time and activating only relevant bus segments. The
segmentation would also pave the way for parallel and independent coupling
calculations, and speeding up the dynamics. These additions would not
significantly affect the area-per-spin, which is dominated by the area of
the state-capacitors.

\section{Conclusion}

The Ising model of computation reformulates computing tasks as set
partitioning problems and, subsequently, as the Ising model ground state
problem. It presents a promising way to find and implement efficient heuristics to
NP-hard combinatorial optimization problems. To compete with solving these
problems on general-purpose computers, however, the prospective
architectures based on the Ising model of computation must support a large
number, billions and more, of spins. Achieving such a high degree of
integration encounters not only technological but also fundamental
challenges. For example, the underlying dynamical model must be, on the one
hand, sufficiently simple to reduce sensitivity to unavoidable variations
of components characteristics. On the other hand, the model should not
sacrifice computational capabilities. To tackle the problems posed by large
architectures, the present paper explores \emph{engineering} the dynamical
model governing the Ising machine.

We show that Ising machines based on Kuramoto networks of nonlinear oscillators
implement the dynamics of the XY model, which, in turn, corresponds to the
gradient-descent solution of the rank-\(2\) semidefinite programming (SDP)
relaxation of the Ising model. This chain of equivalences puts developing Ising
machines into the context of the combinatorial optimization theory and highlights
the dynamical elements enabling the computational resource.

Building on these findings, we consider an approximation of the XY model by
unconstrained dynamical variables with piece-wise linear coupling
(almost-linear Ising machine). We show that the introduced dynamical model,
which we call the triangular model, is characterized by the integrality gap
close to the canonical Goemans-Williamson ratio. Thus, the binary
configuration recovered from the ground state of the triangular model is
close to the ground state of the Ising model. Comparison on a set of
benchmark problems showed that the Ising solver based on the triangular
model provides solutions with the quality on par with \texttt{Circut}, one
of the best max-cut heuristic solvers~\cite{dunningWhat2018}, even when the
machine runs unsupervised (without any preprocessing and intermediate
controls).

We study the scaling properties of the almost-linear Ising machine on an
ensemble of random Erd\H{o}s-R\'{e}nyi graphs $\sset{G}_{N,p}$ with $200 \leq N
\leq 4000$ and $0.05 \leq p \leq 0.35$ and compare its performance with
\texttt{Circut}. We show that the running time of the machine scales
polynomially with the size of the problem. On moderately sized problems (with
the number of edges exceeding $10^5$), the almost-linear Ising machine
showed orders of magnitude speed up compared to \texttt{Circut} with below
one percent solution discrepancy.

To demonstrate opportunities provided by the almost-linear dynamical models
for hardware realizations, we investigate a proof-of-concept circuit
implementation and verify it using the device-level simulation. Thanks to
the simplicity of the dynamics, the designed architecture demonstrates
small area per spin, increased state programmability, and flexibility to
simulate all-to-all connected graphs.

\bmhead{Acknowledgments}

The work has been supported by the US National Science Foundation (NSF) under
Grant No. 1710940 and by the US Air Force Office of Scientific Research
(AFOSR) under Grant No. FA9550-16-1-0363.



\end{document}